\documentclass[a4paper]{article}
\usepackage{graphicx, cite}
\usepackage{multirow}
\usepackage{INTERSPEECH2019}
\usepackage{subfigure}
\usepackage{url}

\title{Acoustic Modeling for Automatic Lyrics-to-Audio Alignment}
\name{Chitralekha Gupta, Emre Y{\i}lmaz, Haizhou Li}
\address{
  Department of Electrical and Computer Engineering, National University of Singapore}
\email{\{chitralekha, emre, haizhou.li\}@nus.edu.sg}

\begin{document}

\maketitle
\begin{abstract}
 Automatic lyrics to polyphonic audio alignment is a challenging task not only because the vocals are corrupted by background music, but also there is a lack of annotated polyphonic corpus for effective acoustic modeling. In this work, we propose (1) using additional speech and music-informed features and (2) adapting the acoustic models trained on a large amount of solo singing vocals towards polyphonic music using a small amount of in-domain data. Incorporating additional information such as voicing and auditory features together with conventional acoustic features aims to bring robustness against the increased spectro-temporal variations in singing vocals. By adapting the acoustic model using a small amount of polyphonic audio data, we reduce the domain mismatch between training and testing data. We perform several alignment experiments and present an in-depth alignment error analysis on acoustic features, and model adaptation techniques. The results demonstrate that the proposed strategy provides a significant error reduction of word boundary alignment over comparable existing systems, especially on more challenging polyphonic data with long-duration musical interludes.

\end{abstract}
\vspace{-0.2cm}
\section{Introduction}
\vspace{-0.15cm}
The goal of an automatic lyrics-to-audio alignment algorithm is the time synchronization between the lyrics and the singing vocals with or without background music. It potentially enables various applications such as generating karaoke scrolling lyrics, music video subtitling, and music retrieval.  

The task of lyrics-to-audio alignment is often seen as an extension of the speech-to-text alignment task. ASR systems have been used to force-align lyrics to singing vocals \cite{fujihara2011lyricsynchronizer,mauch2012integrating,mcvicar2014leveraging,mesaros2010automatic,guptasemi}. Singing voice, however, covers a much wider range of intrinsic variations than speech both in terms of timbre and fundamental frequencies \cite{ramona2008}. One can reduce the mismatch between speech and singing signals by adapting the speech acoustic models with a small amount of singing data using maximum a posterior (MAP) or maximum likelihood linear regression (MLLR) \cite{guptasemi,mesaros2010automatic}. Mesaros et al.~\cite{mesaros2010automatic} used 49 fragments of songs, 20-30 seconds long, along with their manual transcriptions to adapt Gaussian mixture model (GMM)-hidden Markov model (HMM) speech models for singing. These studies provide a direction for solving the problem of lyrics alignment in music, but they suffer from a lack of lyrics annotated data.

Kruspe \cite{kruspe2016bootstrapping} and Dzhambazov \cite{dzhambazov2015modeling} presented systems for the lyrics alignment challenge in MIREX 2017. The acoustic models in \cite{kruspe2016bootstrapping} were trained using 6,000 songs from the Smule's public solo-singing karaoke dataset called Digital Archive of Mobile Performances (DAMP)~\cite{DAMPData}. This dataset is collected via a karaoke app, therefore has no consistent recording condition, contains out-of-vocabulary words, and incorrectly pronounced words because of unfamiliar lyrics\cite{guptasemi}. Moreover, the dataset does not have lyrics time annotation.

Gupta et al.~\cite{guptasemi} designed a semi-supervised algorithm to automatically obtain weak line-level lyrics annotation of a subset of approximately 50 hours of solo-singing DAMP data. They adapted DNN-HMM speech acoustic models to singing voice with this data, that showed 36.32\% word error rate (WER) in a free-decoding experiment on short solo-singing test phrases from the same dataset. In \cite{gupta2018automatic}, these singing-adapted models were further enhanced to capture long duration vowels with a duration-based lexicon modification, that reduced the WER to 29.65\%. However, acoustic models trained on solo-singing data result in a significant drop in performance when applied to singing vocals in the presence of background music\footnote{\url{https://www.music-ir.org/mirex/wiki/2017:Automatic_Lyrics-to-Audio_Alignment_Results}}. Singing vocals are often highly correlated with the corresponding background music, resulting in overlapping frequency components \cite{ramona2008}. The varied range of voice quality of artists combined with different types of musical instruments makes the problem of lyrics alignment highly challenging in polyphonic music.

To suppress the background accompaniment, some approaches have incorporated singing voice separation techniques as a pre-processing step \cite{gupta2019,mesaros2010automatic,dzhambazov2015modeling,fujihara2011lyricsynchronizer}. However, this step makes the system dependent on the performance of the singing voice separation algorithm, as the separation artifacts may make the words unrecognizable. Moreover, this requires a separate training setup for the singing voice separation system.

Recently, multiple researchers have explored data intensive approaches to lyrics-to-audio alignment. In MIREX 2018, Wang \cite{wang2018mirex} presented a system that achieved a mean alignment error (AE) of 4.12 seconds on a standard test data for word alignment evaluation (Mauch's polyphonic dataset \cite{mauch2012integrating}). They used 7,300 annotated English songs from KKBOX Inc.'s music library to train GMM-HMM models. Stoller et al.~\cite{stoller2019} presented an end-to-end system based on the Wave-U-Net architecture that predicts character probabilities directly from raw audio. The system was trained on more than 44,000 songs with line-level lyrics annotations from the Spotify's music library. They achieved an impressive 0.35s mean AE on the Mauch's dataset. However, end-to-end systems require a large amount of annotated training polyphonic music data to perform well as seen in \cite{stoller2019}, while publicly available acoustic resources for polyphonic music are limited.

In this study, we explore the use of additional speech and music-informed features, along with the standard acoustic features during the acoustic model training for singing voice. In addition, we adapt an acoustic model trained on a large amount of solo singing vocals using a limited amount of annotated polyphonic data to reduce the domain mismatch. The aim is to investigate the performance of content-informed features and adaptation methods in capturing the spectro-temporal characteristics of singing voice in polyphonic music.

\vspace{-0.2cm}
\section{Speech and music-informed features}
\label{sec:2}
\vspace{-0.15cm}
Speech and singing have many similarities because they share the underlying physiological mechanisms for production, such as articulatory movements in vocal production \cite{zatorre2012musical,zhang2014study}. Modern ASR systems use conventional acoustic features such as mel-scaled cepstral coefficients (MFCC) to capture the phonetic aspects in conjunction with speaker representations such as i-vectors~\cite{dehak2011} to capture speaker information. These features have been widely used for various MIR tasks such as genre classification, artist, and song identification \cite{tzanetakis2002musical,park2018hybrid,mandel2005}. However, the acoustic characteristics of singing and speech also differ in many ways, such as pitch range, vibrato, and phoneme duration \cite{fujihara2012lyrics, loscos1999low}. Moreover, the presence of different kinds of musical accompaniments, along with singing vocals, constitute additional frequency components in the music signal, that may render the lyrics unrecognizable \cite{ramona2008}. We hypothesize that including additional speech and music informed low-level descriptors for acoustic modeling of sung lyrics will result in improved lyrics-to-audio alignment. Low-level descriptors provide discriminatory information about the temporal variations of the background music and the transitions between sung phonemes and notes, in addition to the timbral information provided by the conventional MFCC features.

The open-source feature extractor called OpenSMILE (or Open Speech and Music Interpretation by Large-space Extraction) \cite{eyben2010opensmile} unites feature extraction algorithms from the speech processing and the MIR communities. It provides various audio low-level descriptors (LLD) that have been widely used for emotion recognition in speech \cite{schuller2013interspeech}, as well as for summarization \cite{raposo2019information}, mood classification \cite{alajanki2016benchmarking}, and singing quality assessment \cite{bohm2017seeking} in music. In this work, we have divided these features into five feature groups, namely \textit{voicing (V), energy (E), auditory (A), spectral (S), and chroma (C)}, as described in Table \ref{tab:lld}.

As indicated in early studies in speech-music discrimination \cite{carey1999comparison,panagiotakis2005speech}, the distribution of the first differential of pitch in singing voice shows a high concentration around zero delta pitch corresponding to steady notes. A similar behavior is observed for the delta amplitude as well. Also, large changes in pitch are observed in singing corresponding to transition between notes. These aspects are covered by the \textit{voicing} and \textit{energy} feature groups.

Singing vocals in presence of background music or chorus is similar to speech in the presence of noise. Relative spectra (RASTA) \cite{hermansky1994rasta} is a filtered representation of an audio signal that is robust to additive and convolutional noise. It essentially suppresses the spectral components that change more quickly or slowly compared to the typical range of speaking rate. Therefore, the \textit{auditory} feature group is expected to be robust to background music and chorus.

\textit{Spectral} group of features represent the ``musical surface'' which denote the characteristics of music related to texture, timbre and instrumentation, as coined by Tzanetakis et al.~\cite{george2001automatic}. The statistics of the distribution of various spectral descriptors such as spectral centroid, flux, energy over time represent the musical surface for pattern recognition purposes.

\textit{Chroma} features have been used previously for tasks such as cover song identification, and music audio classification \cite{ellis2007}. These features consist of a 12-element vector with each dimension representing the intensity associated with a particular musical semitone. While spectral features such as MFCCs represent the timbral characteristics, chroma features reflect the harmonic and melodic content of the music signal, and are shown to provide information independent of the spectral features \cite{ellis2007}.

\begin{table}[]
\centering
\caption{Description of 5 acoustic feature groups.}
\label{tab:lld}
\vspace{-0.3cm}
\resizebox{\columnwidth}{!}{%
\begin{tabular}{cccc}
\hline
\begin{tabular}[c]{@{}c@{}}\textbf{Group}\\\textbf{ID}\end{tabular} & \begin{tabular}[c]{@{}c@{}}\textbf{Feature}\\\textbf{Group}\end{tabular} & \textbf{Description} & \textbf{\#LLDs} \\ \hline
A & Auditory & \begin{tabular}[c]{@{}c@{}}RASTA-style auditory spectrum\\ bands 1-26 (0-8 kHz)\end{tabular} & 26 + deltas \\ \hline
E & Energy & \begin{tabular}[c]{@{}c@{}}Sum of auditory spectrum (loudness), \\ sum of RASTA-style auditory spectrum, \\ RMS energy, zero crossing rate\end{tabular} & 4 + deltas \\ \hline
C & Chroma & Intensities in 12 musical semitones & 12 \\ \hline
S & Spectral & \begin{tabular}[c]{@{}c@{}}Spectral energy 250-650 Hz, 1 k-4kHz\\ Spectral Roll Off Point 0.25, 0.50, 0.75, 0.90\\ Spectral Flux, Entropy, Variance, Skewness, Kurtosis,\\ Slope, Psychoacoustic Sharpness, Harmonicity\end{tabular} & 15 + deltas \\ \hline
V & Voicing & \begin{tabular}[c]{@{}c@{}}F0, Voicing, Jitter (local, delta),\\ Shimmer, Logarithmic HNR\end{tabular} & 6 + deltas \\ \hline
\end{tabular}%
}
\vspace{-0.5cm}
\end{table}
\vspace{-0.3cm}
\section{Model adaptation for domain mismatch}
\label{sec:3}
\vspace{-0.15cm}
Our goal is to build a framework to automatically align lyrics to the polyphonic music audio. With an acoustic model trained on solo-singing data, we can adapt the model towards the test data in two ways: (a) by making the test data closer to the trained solo-singing acoustic models by applying vocal separation on polyphonic test data, (b) by adapting the acoustic models to polyphonic data. In \cite{gupta2019}, the former approach was explored. But source separation algorithms are known to introduce artifacts in the extracted vocal, thus the pipeline gets dependent on the reliability of the source separation algorithm. In this work, we investigate the latter approach, i.e.~adapting the acoustic model using a small amount of in-domain polyphonic data to reduce the domain mismatch. Model adaptation is achieved by initializing the hidden layers using the neural network acoustic model trained on the solo-singing data and retraining this model by performing extra forward-backward passes only using the available polyphonic training data for a small number of epochs and possibly with a smaller learning rate.

As discussed earlier, acoustic modeling of singing vocals in the presence of background music is constrained by a lack of lyrics annotated data. Recently, a multimodal DALI dataset \cite{meseguer2018dali} was introduced, that consists of 5,000+ polyphonic songs with note annotations and weak word-level, line-level, and paragraph-level lyrics annotations. It was created with a set of initial manual annotations of time-aligned lyrics made by non-expert users of Karaoke games, where the audio was not available. The corresponding audio candidates were then retrieved from the web, and an iterative method of obtaining a large-scale lyrics annotated polyphonic music data was proposed. However, the reliability of these lyrics annotations have not been verified. The authors have released 105 songs as the ground-truth data, where the annotations are manually checked and corrected. In this work, we make use of this ground-truth data for domain adaptation.
\vspace{-0.3cm}
\section{Experimental setup}
\label{sec:4}
\vspace{-0.15cm}
We conduct two sets of experiments to study the impact of our proposed acoustic modeling strategies for lyrics alignment: (1) we first assess the effect of the speech and music informed features on lyrics alignment in solo-singing, and (2) then we investigate the effects of these features in polyphonic music lyrics alignment, along with model adaptation techniques. In this section, we detail the datasets used for the experiments, acoustic model architecture, the system configurations, and evaluation metrics for assessing the quality of the boundaries.
\begin{table}[]
\centering
\caption{Dataset description. (solo: solo-singing; poly: singing mixed with music)}
\label{tab:datasets}
\vspace{-0.3cm}
\resizebox{\columnwidth}{!}{%
\setlength\tabcolsep{1.5pt}
\begin{tabular}{c|c|c|c|c}
\hline
\textbf{Name} & \begin{tabular}[c]{@{}c@{}}\textbf{Audio}\\\textbf{type} \end{tabular}& \textbf{Content} & \textbf{Lyrics Ground-Truth} & \begin{tabular}[c]{@{}c@{}}\textbf{Avg Word}\\ \textbf{Length(s)/\# words}\end{tabular}\\ \hline
\multicolumn{5}{c}{\textbf{Training/Adaptation data}} \\ \hline
DAMP train \cite{guptasemi} & solo & \begin{tabular}[c]{@{}c@{}}35,662 lines\end{tabular} & line-level weak transcription & -\\ 
DALI train \cite{meseguer2018dali} & poly & \begin{tabular}[c]{@{}c@{}}70 songs\end{tabular}& word and line-level boundaries &\\ \hline 
\multicolumn{5}{c}{\textbf{Test data}} \\ \hline
DAMP test \cite{guptasemi} & solo & \begin{tabular}[c]{@{}c@{}}1697 lines\end{tabular} & line-level transcription & -\\ 
Hansen-solo \cite{hansen2012recognition} & solo & 7 songs & word-level boundaries & 0.485 / 2212\\ 
Hansen-poly \cite{hansen2012recognition} & poly & 7 songs & word-level boundaries & 0.485 / 2212\\ 
Mauch-poly \cite{mauch2012integrating} & poly & 20 songs & word-level boundaries & 0.871 / 5052\\
DALI dev \cite{meseguer2018dali}&poly&9 songs&word and line-level boundaries & 0.471 / 2305\\
DALI test \cite{meseguer2018dali}&poly&20 songs&word and line-level boundaries & 0.442 / 5260 \\\hline
\end{tabular}%
}
\vspace{-0.1cm}
\end{table}
\vspace{-0.2cm}
\subsection{Datasets}
\label{ssec:4.1}
\vspace{-0.2cm}
All datasets used in the experiments are summarized in Table \ref{tab:datasets}. The training data for solo-singing acoustic modeling is approximately 50 hours of the DAMP dataset \cite{DAMPData,guptasemi} that has weak line-level lyrics transcription. We use the DALI ground-truth data for domain adaptation of the acoustic models to the polyphonic music. It consists of 99 songs\footnote{There are a total of 105 songs in the ground-truth data, out of which the audio file links to 6 songs are not accessible from Singapore.}, that we divided into train, development (dev), and test, in the ratio of 70:9:20.

We evaluated our alignment systems on two datasets - 7 songs\footnote{The word boundary ground-truth of the songs \textit{clocks} and \textit{i kissed a girl} were not accurate, hence excluded from this study} from the Hansen's a capela and polyphonic datasets \cite{hansen2012recognition}, and 20 songs of the Mauch's polyphonic dataset \cite{mauch2012integrating}. These datasets were used in the MIREX lyrics alignment challenges of 2017 and 2018. These datasets consist of Western pop songs with manually annotated word-level boundaries. We tune our model adaptation scheme on the DALI-dev set, and also report alignment results on the DALI-test set.

\begin{table}[]
\centering
\caption{System configurations. Baseline acoustic models are trained on DAMP subset-train (Table \ref{tab:datasets}). AECSV are the feature group IDs from Table \ref{tab:lld}. }
\label{tab:configs}
\vspace{-0.3cm}
\resizebox{0.75\columnwidth}{!}{%
\begin{tabular}{c|c|c}
\hline
\textbf{Configs} & \textbf{Adaptation data} & \textbf{Features}\\ \hline
C1 & - & MFCC, i-vectors \\ 
C2 & - & MFCC, i-vectors, AECSV\\ 
C3 & vocal-extracted DALI & MFCC, i-vectors\\ 
C4 & vocal-extracted DALI& MFCC, i-vectors, AECSV\\ 
C5 & polyphonic DALI & MFCC, i-vectors\\ 
C6 & polyphonic DALI& MFCC, i-vectors, AECSV\\ \hline
\end{tabular}%
}
\vspace{-0.5cm}
\end{table}
\vspace{-0.2cm}
\subsection{ASR architecture}
\vspace{-0.2cm}
The ASR system used in these experiments is trained using the Kaldi ASR toolkit~\cite{povey2011kaldi}. A context dependent GMM-HMM system is trained with 40k Gaussians using 39 dimensional MFCC features including the deltas and delta-deltas to obtain the alignments for neural network training. The frame rate and length are 10 and 25 ms, respectively. A factorized time-delay neural network (TDNN-F) model~\cite{povey2018} with additional convolutional layers (2 convolutional, 10 time-delay layers followed by a rank reduction layer) was trained according the standard Kaldi recipe (version 5.4). An augmented version of the solo-singing training data described in Section~\ref{ssec:4.1} is created by reducing (x0.9) and increasing (x1.1) the speed of each utterance~\cite{ko2015}. This augmented training data is used for training the neural network-based acoustic model. The default hyperparameters provided in the standard recipe were used and no hyperparameter tuning was performed during the acoustic model training. The baseline acoustic model is trained using 40-dimensional MFCCs as acoustic features that are combined with i-vectors~\cite{saon2013}. During the training of the neural network~\cite{povey2016}, the frame subsampling rate is set to 3 providing an effective frame shift of 30 ms. A duration-based modified pronunciation lexicon is employed which is detailed in~\cite{gupta2018automatic}.
\vspace{-0.2cm}
\subsection{System configurations}
\vspace{-0.15cm}
The baseline acoustic model (C1) is trained on solo-singing DAMP subset-train with the 40-dimensional MFCCs and 100-dimensional i-vectors. To test the performance of the additional features, extracted using OpenSMILE toolbox \cite{eyben2010opensmile}, we append the five feature groups with a total dimension of 154 to the 140-dimensional baseline feature vector (C2). We also analyse the contribution of each feature group by appending only one feature subset, eg. C2-V, C2-A, C2-E etc. We adapt the baseline model with the vocal-extracted DALI-train data (C3, C4), and polyphonic DALI-train data (C5, C6). We use the state-of-the-art implementation of the Wave-U-Net based audio source separation \cite{stoller2018} for vocal extraction from the polyphonic audio.
\vspace{-0.2cm}
\subsection{Evaluation metrics}
\vspace{-0.15cm}
Mean AE is the absolute error or deviation in seconds from the predicted to the true word start times, averaged over all words in a dataset. Previous studies have reported this metric, but mean AE is affected drastically by outliers. Therefore, to gauge the distribution of alignment errors, we also present median (Med.), standard deviation (Std.) of the absolute boundary errors. Moreover, we measure the percentage of hypothesized word boundaries that are within an acceptable tolerance interval around the ground-truth boundary  (i.e.~\%Correct or \%C). Observing the range of average word durations in Table \ref{tab:datasets}, we set this acceptable tolerance interval as approximately half the average duration of words, i.e.~the percentage of word-start boundaries within 250 ms of the ground-truth.
\begin{table}[th]
  \caption{Mean AE performance on Hansen's solo-singing data with models trained on DAMP solo-singing data. The median of absolute boundary errors in all cases in this table is 0.03s.}
  \label{tab:solosinging}
  \vspace{-0.3cm}
  \centering
  \resizebox{0.35\columnwidth}{!}{
  \setlength\tabcolsep{1.5pt}
  \begin{tabular}{c c c c}
    \toprule
    \textbf{Config} & \textbf{Mean(s)} & \textbf{Std.(s)}& \textbf{\%C}\\
    \midrule
    C1                       & 0.20 & 0.75& 91.5           \\
    C2                      & 0.13 & 0.63& 94.1     \\
    C2-A                      & 0.17    & 0.95& 92.3  \\
    C2-E                      & 0.30   & 1.73&91.7    \\
    C2-C                     & 0.32   & 1.84&90.7    \\
    C2-S                      & 0.24   & 1.36 & 92.7   \\
    C2-V                      & 0.48  & 1.75&87.6     \\
    \bottomrule
  \end{tabular}
  }
\end{table}

\begin{table}[]
\centering
\caption{Mean AE for various adaptation configurations (LR: same initial learning rate; 0.5LR: half of initial learning rate).}
\label{tab:adapt}
\vspace{-0.3cm}
\resizebox{\columnwidth}{!}{%
\begin{tabular}{c|c|c|c|c|c|c|c}
\hline
 \textbf{Config -\textgreater{}} & \textbf{C1} & \textbf{\begin{tabular}[c]{@{}c@{}}LR,\\ epoch1\end{tabular}} & \textbf{\begin{tabular}[c]{@{}c@{}}LR, \\ epoch2\end{tabular}} & \textbf{\begin{tabular}[c]{@{}c@{}}LR, \\ epoch3\end{tabular}} & \textbf{\begin{tabular}[c]{@{}c@{}}0.5LR,\\ epoch1\end{tabular}} & \textbf{\begin{tabular}[c]{@{}c@{}}0.5LR,\\ epoch2\end{tabular}} & \textbf{\begin{tabular}[c]{@{}c@{}}0.5LR,\\ epoch3\end{tabular}} \\ \hline
DALI-dev & 0.288 & \textbf{0.170} & 0.182 & 0.173 & 0.171 & 0.198 & 0.201 \\ 
DALI-test & 0.343 & 0.159 & 0.162 & 0.163 & 0.156 & 0.176 & 0.174 \\ \hline
\end{tabular}%
}
\end{table}
\vspace{-0.5cm}
\section{Results and discussion}
\label{sec:5}
\vspace{-0.15cm}
\subsection{Performance on solo-singing}
\vspace{-0.15cm}
In the first set of experiments, we explore the effect of each of the speech and music informed feature groups combined with MFCCs and i-vectors. The alignment results provided by different feature configurations on the Hansen's solo-singing dataset is shown in Table \ref{tab:solosinging}. Training the solo-singing acoustic models with the additional features reduces the average boundary error from 200 ms to 130 ms, while the standard deviation and the \%C also improve. 
We also observe that the auditory and the spectral feature groups individually contribute to the improved performance. Many songs in this solo-singing dataset contain chorus sections, where other singers and the main singer may sing different lyrics at the same time. The robust RASTA features in auditory group is observed to be helpful in such cases. Moreover, the individual groups perform worse than their combination, which implies that the groups provide exclusive information that complement each other.
\begin{table}[]
\centering
\caption{AE performance on \textbf{vocal-extracted} Hansen-poly and Mauch-poly data.}
\label{tab:vocal_extract}
\vspace{-0.3cm}
\resizebox{0.8\columnwidth}{!}{%
\setlength\tabcolsep{1.5pt}
\begin{tabular}{c||c|c|c|c||c|c|c|c}
\hline
 & \multicolumn{4}{c||}{\textbf{Hansen-poly}} & \multicolumn{4}{c}{\textbf{Mauch-poly}} \\ \hline
 & \textbf{Med.(s)} & \textbf{Mean(s)} & \textbf{Std.(s)} & \textbf{\%C} & \textbf{Med.(s)} & \textbf{Mean(s)} &\textbf{ Std.(s)} & \textbf{\%C} \\ \hline
C1 & 0.23 & 2.33 & 5.10 & 51.4 & 1.49 & 14.31 & 22.37 & 32.8 \\ 
C2 & 0.15 & 0.94 & 2.76 & 69.9 & 0.26 & 4.05 & 8.30 & 49.0 \\ 
C3 & 0.82 & 6.84 & 11.92 & 41.1 & 1.61 & 12.47 & 20.39 & 34.9 \\ 
C4 & 0.21 & 2.35 & 5.22 & 59.6 & 0.36 & 5.19 & 9.52 & 41.8 \\ \hline
\end{tabular}%
}
\vspace{-0.1cm}
\end{table}
\begin{table}[]
\centering
\caption{AE performance on Hansen-poly and Mauch-poly data.}
\label{tab:poly}
\vspace{-0.3cm}
\resizebox{0.8\columnwidth}{!}{%
\setlength\tabcolsep{1.5pt}
\begin{tabular}{c||c|c|c|c||c|c|c|c}
\hline
\multirow{2}{*}{} & \multicolumn{4}{c||}{\textbf{Hansen-poly}} & \multicolumn{4}{c}{\textbf{Mauch-poly}} \\ \cline{2-9} 
 & \textbf{Med.(s)} & \textbf{Mean(s)} & \textbf{Std.(s)} & \textbf{\%C} & \textbf{Med.(s)} & \textbf{Mean(s)} &\textbf{ Std.(s)} & \textbf{\%C} \\ \hline
C1 & 30.10 & 36.20 & 31.85 & 14.5 & 20.33 & 39.70 & 48.55 & 10.5 \\ 
C2 & 2.88 & 9.57 & 13.38 & 27.7 & 2.93 & 14.69 & 22.59 & 25.8 \\
C5 & 0.08 & 1.82 & 5.72 & 71.8 & 0.15 & 3.78 & 9.98 & 60.9 \\ 
C6 & 0.11 & 2.37 & 6.85 & 64.7 & 0.18 & \textbf{1.93} & 5.90 & 57.5 \\ \hline
\end{tabular}%
}
\vspace{-0.1cm}
\end{table}
\vspace{-0.2cm}
\subsection{Performance on polyphonic audio}
\vspace{-0.15cm}
To reduce the domain mismatch between solo-singing acoustic models and the polyphonic test data, we adopt three approaches: (a) vocal extraction of the polyphonic test data, as done in previous studies \cite{gupta2019,mesaros2010automatic,kruspe2016bootstrapping}, (b) adapt the models with vocal extracted polyphonic data, and (c) adapt the models with polyphonic data. We used DALI-train for adaptation, and DALI-dev to optimize the alignment performance (mean AE) by adjusting the initial learning rate (LR) and the number of epochs, as shown in Table \ref{tab:adapt}. We choose the setting that performs the adaptation using the same initial LR within a single epoch as it gives the best performance on the development set. The best result reported on the DALI-test set is also obtained using this setting. Please note that the DALI-test data contains short lines or utterances of 3-10s, which is different from the other test sets in which the entire song of 2-3 mins. is given to the system. The short duration of the DALI-test set results in relatively smaller mean AE values.
\vspace{-0.3cm}
\subsubsection{On vocal-extracted polyphonic test data}
\vspace{-0.15cm}
Table \ref{tab:vocal_extract} summarizes the performance of solo-singing models (C1, C2) and adapted models with extracted vocals (C3, C4) with and without the additional features on the vocal extracted Hansen-poly and Mauch-poly test datasets. We observe that model adaptation does only a slight difference in the performance (cf. C1, C3), but the additional features improve the performance by a large margin (cf. C1, C2). MFCC features are known to be sensitive to background noise \cite{li2014}. So, domain adaptation with the extracted vocals containing distortion and artifacts is a possible reason for the poor performance of the adapted models. On the other hand, the additional features are designed to be robust to noise, thus improving the performance.
\vspace{-0.58cm}
\subsubsection{On polyphonic test data (without vocal extraction)}
\vspace{-0.15cm}
Table \ref{tab:poly} shows the lyrics alignment performance of the unadapted (C1, C2) and polyphonic data adapted (C5, C6) acoustic models on the Hansen-poly and Mauch-poly data. The poor performance of the solo-singing models (C1, C2) on polyphonic data is expected due to domain mismatch. But here, the domain adaptation (C5, C6) gives a considerable improvement in performance. A comparison of Table \ref{tab:vocal_extract} and \ref{tab:poly} shows that domain adaptation \textit{without} vocal extraction performs better. This suggests that domain adaptation with a small amount of polyphonic data helps the acoustic model capture the spectro-temporal variations of singing vocals, which offers a simple, but effective solution in scenarios with limited polyphonic singing data. 

One main difference between the Hansen's and Mauch's datasets is that the songs in the Mauch's dataset are rich in long-duration musical interludes that have no singing vocals, while Hansen's has only a few of such interludes. We observe that the content-informed features and domain adaptation help to improve the boundaries next to these long interludes. Thus, the improvement in alignment performance is more evident in the Mauch's dataset, than in the Hansen's dataset.

\begin{figure}
\centering
\includegraphics[width=0.65\columnwidth]{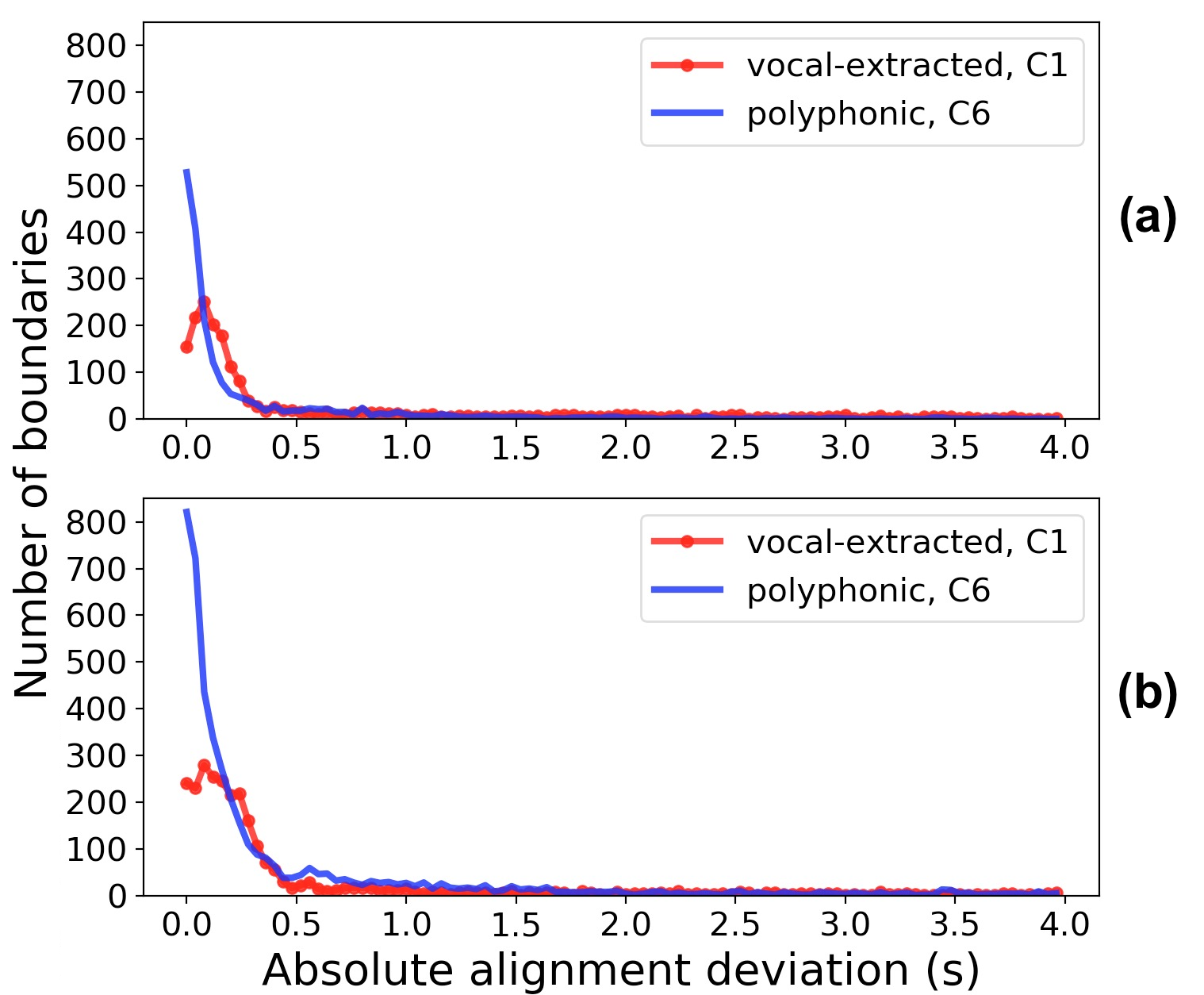}
\vspace{-0.3cm}
\caption{Comparison of word boundary alignment error distribution between C1 on extracted vocals test data and C6 on polyphonic test data on (a) Hansen's and (b) Mauch's datasets.}
\label{fig:hist}
\end{figure}

Although the mean AE of the boundaries is more than a second, the median of errors is less than 180 ms for the best performing systems. A comparison of the boundary error distribution of C1 on extracted vocals, and C6 on polyphonic data (Figure \ref{fig:hist}) shows a large increase in the number of boundaries towards zero error, for both the datasets. This also means that the there are hypothesized boundaries that are far away from the the true boundaries, that needs to be investigated in future.
\vspace{-0.2cm}
\subsection{Comparison with existing literature}
\vspace{-0.15cm}
In Table \ref{tab:litreview}, we compare our best results with the past studies, and find that our strategy provides better results than all previous work, except for the end-to-end system\cite{stoller2019}. An end-to-end system requires a large amount of data for reliable output which we do not have access to. Our proposed strategies show a way to fuse knowledge-driven and data-driven methods to address the problem of lyrics-to-audio alignment in a low-resourced setting.
\begin{table}[]
\centering
\caption{Comparison of mean AE (s) with existing literature.}
\label{tab:litreview}
\vspace{-0.3cm}
\resizebox{\columnwidth}{!}{%
\setlength\tabcolsep{1.5pt}
\begin{tabular}{c|c|c|c|c|c|c}
\hline
\textbf{} & \multicolumn{2}{c|}{\textbf{MIREX 2017}} & \textbf{MIREX 2018} & \multicolumn{2}{c|}{\textbf{ICASSP 2019}} & \textbf{} \\ \hline
\textbf{} & \textbf{AK\cite{kruspe2016bootstrapping}} & \textbf{GD\cite{dzhambazov2015modeling,dzhambazov2017knowknowledge}} & \textbf{CW \cite{wang2018mirex}} & \textbf{DS\cite{stoller2019}} & \textbf{CG\cite{gupta2019}} & \textbf{Ours} \\ \hline
\textbf{Training data} & \begin{tabular}[c]{@{}c@{}}6,000 songs\\ (DAMP)\\(solo)\end{tabular} & \begin{tabular}[c]{@{}c@{}}6,000 songs\\  (DAMP)\\(solo)\end{tabular} & \begin{tabular}[c]{@{}c@{}}7,300 songs\\(KKBOX)\\(poly)\end{tabular} & \begin{tabular}[c]{@{}c@{}}44,232 songs\\ (Spotify)\\(poly)\end{tabular} & \begin{tabular}[c]{@{}c@{}}35,662 lines\\(DAMP)\\(solo)\end{tabular} & \begin{tabular}[c]{@{}c@{}}35,662 lines\\(DAMP) (solo)\\+ 70 songs (DALI)\\(poly)\end{tabular} \\
\textbf{Architecture} & DNN-HMM & DNN-HMM & GMM-HMM & \begin{tabular}[c]{@{}c@{}}UNet based\\ end-to-end\end{tabular} & \begin{tabular}[c]{@{}c@{}}SAT\\DNN-HMM \end{tabular}& CNN-TDNN-F \\\hline
\textbf{Hansen-poly} & 7.34 & 10.57 & 2.07 & - & 1.39 & \begin{tabular}[c]{@{}c@{}}0.93\\(median: 0.15)\end{tabular} \\ 
\textbf{Mauch-poly} & 9.03 & 11.64 & 4.13 & 0.35 & 6.34 & \begin{tabular}[c]{@{}c@{}}1.93\\(median: 0.18)\end{tabular} \\ \hline
\end{tabular}%
}
\vspace{-0.5cm}
\end{table}

\vspace{-0.4cm}
\section{Conclusions}
\vspace{-0.15cm}
In this study, we discuss two strategies to obtain improved acoustic modeling for the task of lyrics-to-audio alignment. Particularly, we propose to (1) employ additional features with speech- and music-related information together with conventional MFCCs, and (2) adapt solo-singing acoustic model using small amount of in-domain polyphonic data. We validated the robustness of these features to background music and ability to capture the spectro-temporal variations in polyphonic singing vocals. The alignment experiments demonstrate that applying the described strategies reduces the mean AE to 1.9s on the Mauch's dataset which is better than all results reported in the MIREX lyrics alignment challenge.
\section{Acknowledgments}
This research is supported by Ministry of Education, Singapore AcRF Tier 1 NUS Start-up Grant FY2016, Non-parametric approach to voice morphing.

\bibliographystyle{IEEEtran}

\bibliography{mybib}

\begin{thebibliography}{10}
\providecommand{\url}[1]{#1}
\csname url@samestyle\endcsname
\providecommand{\newblock}{\relax}
\providecommand{\bibinfo}[2]{#2}
\providecommand{\BIBentrySTDinterwordspacing}{\spaceskip=0pt\relax}
\providecommand{\BIBentryALTinterwordstretchfactor}{4}
\providecommand{\BIBentryALTinterwordspacing}{\spaceskip=\fontdimen2\font plus
\BIBentryALTinterwordstretchfactor\fontdimen3\font minus
  \fontdimen4\font\relax}
\providecommand{\BIBforeignlanguage}[2]{{%
\expandafter\ifx\csname l@#1\endcsname\relax
\typeout{** WARNING: IEEEtran.bst: No hyphenation pattern has been}%
\typeout{** loaded for the language `#1'. Using the pattern for}%
\typeout{** the default language instead.}%
\else
\language=\csname l@#1\endcsname
\fi
#2}}
\providecommand{\BIBdecl}{\relax}
\BIBdecl

\bibitem{fujihara2011lyricsynchronizer}
H.~Fujihara, M.~Goto, J.~Ogata, and H.~G. Okuno, ``Lyricsynchronizer: Automatic
  synchronization system between musical audio signals and lyrics,'' \emph{IEEE
  Journal of Selected Topics in Signal Processing}, vol.~5, no.~6, pp.
  1252--1261, 2011.

\bibitem{mauch2012integrating}
M.~Mauch, H.~Fujihara, and M.~Goto, ``Integrating additional chord information
  into {HMM}-based lyrics-to-audio alignment,'' \emph{IEEE Transactions on
  Audio, Speech and Language Processing}, vol.~20, no.~1, pp. 200--210, 2012.

\bibitem{mcvicar2014leveraging}
M.~McVicar, D.~P. Ellis, and M.~Goto, ``Leveraging repetition for improved
  automatic lyric transcription in popular music,'' in \emph{Proc. ICASSP},
  2014, pp. 3117--3121.

\bibitem{mesaros2010automatic}
A.~Mesaros and T.~Virtanen, ``Automatic recognition of lyrics in singing,''
  \emph{EURASIP Journal on Audio, Speech, and Music Processing}, vol. 2010,
  p.~4, 2010.

\bibitem{guptasemi}
C.~Gupta, R.~Tong, H.~Li, and Y.~Wang, ``Semi-supervised lyrics and
  solo-singing alignment,'' in \emph{Proc. ISMIR}, 2018.

\bibitem{ramona2008}
M.~Ramona, G.~Richard, and B.~David, ``Vocal detection in music with support
  vector machines,'' in \emph{2008 Proc. ICASSP}.\hskip 1em plus 0.5em minus
  0.4em\relax IEEE, 2008, pp. 1885--1888.

\bibitem{kruspe2016bootstrapping}
A.~M. Kruspe, ``Bootstrapping a system for phoneme recognition and keyword
  spotting in unaccompanied singing,'' in \emph{Proc. ISMIR}, 2016, pp.
  358--364.

\bibitem{dzhambazov2015modeling}
G.~B. Dzhambazov and X.~Serra, ``Modeling of phoneme durations for alignment
  between polyphonic audio and lyrics,'' in \emph{12th Sound and Music
  Computing Conference}, 2015, pp. 281--286.

\bibitem{DAMPData}
S.~Sing!, ``Smule.digital archive mobile performances(damp),''
  \url{https://ccrma.stanford.edu/damp/}, 2010 (accessed March 15, 2018).

\bibitem{gupta2018automatic}
C.~Gupta, H.~Li, and Y.~Wang, ``Automatic pronunciation evaluation of
  singing,'' \emph{Proc. INTERSPEECH}, pp. 1507--1511, 2018.

\bibitem{gupta2019}
B.~Sharma, C.~Gupta, H.~Li, and Y.~Wang, ``Automatic lyrics-to-audio alignment
  on polyphonic music using singing-adapted acoustic models,'' in \emph{Proc.
  ICASSP}.\hskip 1em plus 0.5em minus 0.4em\relax IEEE, 2019.

\bibitem{wang2018mirex}
C.-C. Wang, ``Mirex2018: Lyrics-to-audio alignment for instrument accompanied
  singings,'' in \emph{MIREX 2018}, 2018.

\bibitem{stoller2019}
S.~E. Daniel~Stoller, Simon~Durand, ``End-to-end lyrics alignment for
  polyphonic music using an audio-to-character recognition model,'' in
  \emph{Proc. ICASSP}.\hskip 1em plus 0.5em minus 0.4em\relax IEEE, 2019.

\bibitem{zatorre2012musical}
R.~J. Zatorre and S.~R. Baum, ``Musical melody and speech intonation: Singing a
  different tune,'' \emph{PLoS biology}, vol.~10, no.~7, p. e1001372, 2012.

\bibitem{zhang2014study}
S.~Zhang, R.~C. Repetto, and X.~Serra, ``Study of the similarity between
  linguistic tones and melodic pitch contours in {Beijing} opera singing.'' in
  \emph{Proc. ISMIR}, 2014, pp. 343--348.

\bibitem{dehak2011}
N.~Dehak, P.~J. Kenny, R.~Dehak, P.~Dumouchel, and P.~Ouellet, ``Front-end
  factor analysis for speaker verification,'' \emph{IEEE Transactions on Audio,
  Speech, and Language Processing}, vol.~19, no.~4, pp. 788--798, May 2011.

\bibitem{tzanetakis2002musical}
G.~Tzanetakis and P.~Cook, ``Musical genre classification of audio signals,''
  \emph{IEEE Transactions on Speech and Audio Processing}, vol.~10, no.~5, pp.
  293--302, 2002.

\bibitem{park2018hybrid}
J.~Park, D.~Kim, J.~Lee, S.~Kum, and J.~Nam, ``A hybrid of deep audio feature
  and i-vector for artist recognition,'' \emph{arXiv preprint
  arXiv:1807.09208}, 2018.

\bibitem{mandel2005}
M.~Mandel and D.~Ellis, ``Song-level features and support vector machines for
  music classification,'' in \emph{Proc. ISMIR}, 2005.

\bibitem{fujihara2012lyrics}
H.~Fujihara and M.~Goto, ``Lyrics-to-audio alignment and its application,'' in
  \emph{Dagstuhl Follow-Ups}, vol.~3.\hskip 1em plus 0.5em minus 0.4em\relax
  Schloss Dagstuhl-Leibniz-Zentrum fuer Informatik, 2012.

\bibitem{loscos1999low}
A.~Loscos, P.~Cano, and J.~Bonada, ``Low-delay singing voice alignment to
  text.'' in \emph{Proc. ICMC}, 1999.

\bibitem{eyben2010opensmile}
F.~Eyben, M.~W{\"o}llmer, and B.~Schuller, ``Opensmile: the {Munich} versatile
  and fast open-source audio feature extractor,'' in \emph{Proc. ACM
  Multimedia}.\hskip 1em plus 0.5em minus 0.4em\relax ACM, 2010, pp.
  1459--1462.

\bibitem{schuller2013interspeech}
B.~Schuller, S.~Steidl, A.~Batliner, A.~Vinciarelli, K.~Scherer, F.~Ringeval,
  M.~Chetouani, F.~Weninger, F.~Eyben, E.~Marchi \emph{et~al.}, ``The
  interspeech 2013 computational paralinguistics challenge: social signals,
  conflict, emotion, autism,'' in \emph{Proc. INTERSPEECH}, 2013.

\bibitem{raposo2019information}
F.~A. Raposo, D.~M. de~Matos, and R.~Ribeiro, ``An information-theoretic
  approach to machine-oriented music summarization,'' \emph{Pattern Recognition
  Letters}, 2019.

\bibitem{alajanki2016benchmarking}
A.~Alajanki, Y.-H. Yang, and M.~Soleymani, ``Benchmarking music emotion
  recognition systems,'' \emph{PLOS ONE}, pp. 835--838, 2016.

\bibitem{bohm2017seeking}
J.~B{\"o}hm, F.~Eyben, M.~Schmitt, H.~Kosch, and B.~Schuller, ``Seeking the
  superstar: Automatic assessment of perceived singing quality,'' in \emph{2017
  International Joint Conference on Neural Networks (IJCNN)}.\hskip 1em plus
  0.5em minus 0.4em\relax IEEE, 2017, pp. 1560--1569.

\bibitem{carey1999comparison}
M.~J. Carey, E.~S. Parris, and H.~Lloyd-Thomas, ``A comparison of features for
  speech, music discrimination,'' in \emph{Proc. ICASSP}, vol.~1.\hskip 1em
  plus 0.5em minus 0.4em\relax IEEE, 1999, pp. 149--152.

\bibitem{panagiotakis2005speech}
C.~Panagiotakis and G.~Tziritas, ``A speech/music discriminator based on {RMS}
  and zero-crossings,'' \emph{IEEE Transactions on Multimedia}, vol.~7, no.~1,
  pp. 155--166, 2005.

\bibitem{hermansky1994rasta}
H.~Hermansky and N.~Morgan, ``Rasta processing of speech,'' \emph{IEEE
  Transactions on Speech and Audio Processing}, vol.~2, no.~4, pp. 578--589,
  1994.

\bibitem{george2001automatic}
T.~George, E.~Georg, and C.~Perry, ``Automatic musical genre classification of
  audio signals,'' in \emph{Proc. ISMIR}, 2001.

\bibitem{ellis2007}
D.~Ellis, ``Classifying music audio with timbral and chroma features,'' in
  \emph{Proc. ISMIR}, 2007.

\bibitem{meseguer2018dali}
G.~Meseguer-Brocal, A.~Cohen-Hadria, and G.~Peeters, ``Dali: A large dataset of
  synchronized audio, lyrics and notes, automatically created using
  teacher-student machine learning paradigm,'' in \emph{Proc. ISMIR}, 2018.

\bibitem{hansen2012recognition}
J.~K. Hansen, ``Recognition of phonemes in a-cappella recordings using temporal
  patterns and mel frequency cepstral coefficients,'' in \emph{9th Sound and
  Music Computing Conference (SMC)}, 2012, pp. 494--499.

\bibitem{povey2011kaldi}
D.~Povey, A.~Ghoshal, G.~Boulianne, L.~Burget, O.~Glembek, N.~Goel,
  M.~Hannemann, P.~Motlicek, Y.~Qian, P.~Schwarz \emph{et~al.}, ``The {Kaldi}
  speech recognition toolkit,'' in \emph{in Proc. ASRU}, 2011.

\bibitem{povey2018}
D.~Povey, G.~Cheng, Y.~Wang, K.~Li, H.~Xu, M.~Yarmohammadi, and S.~Khudanpur,
  ``Semi-orthogonal low-rank matrix factorization for deep neural networks,''
  in \emph{Proc. INTERSPEECH}, 2018, pp. 3743--3747.

\bibitem{ko2015}
T.~Ko, V.~Peddinti, D.~Povey, and S.~Khudanpur, ``Audio augmentation for speech
  recognition,'' in \emph{Proc. INTERSPEECH}, 2015, pp. 3586--3589.

\bibitem{saon2013}
G.~Saon, H.~Soltau, D.~Nahamoo, and M.~Picheny, ``Speaker adaptation of neural
  network acoustic models using i-vectors,'' in \emph{Proc. ASRU}, Dec 2013,
  pp. 55--59.

\bibitem{povey2016}
D.~Povey, V.~Peddinti, D.~Galvez, P.~Ghahremani, V.~Manohar, X.~Na, Y.~Wang,
  and S.~Khudanpur, ``Purely sequence-trained neural networks for {ASR} based
  on lattice-free {MMI},'' in \emph{Proc. INTERSPEECH}, 2016, pp. 2751--2755.

\bibitem{stoller2018}
D.~Stoller, S.~Ewert, and S.~Dixon, ``Wave-u-net: A multi-scale neural network
  for end-to-end audio source separation,'' in \emph{Proc. ISMIR}, 2018.

\bibitem{li2014}
J.~{Li}, L.~{Deng}, Y.~{Gong}, and R.~{Haeb-Umbach}, ``An overview of
  noise-robust automatic speech recognition,'' \emph{IEEE/ACM Transactions on
  Audio, Speech, and Language Processing}, vol.~22, no.~4, pp. 745--777, April
  2014.

\bibitem{dzhambazov2017knowknowledge}
G.~Dzhambazov, ``Knowledge-based probabilistic modeling for tracking lyrics in
  music audio signals,'' Ph.D. dissertation, Universitat Pompeu Fabra, 2017.

\end{thebibliography}

\end{document}